\documentclass[cits,a4paper]{PoS}
\usepackage[utf8]{inputenc}

\title{The first year of the \textit{Fermi} Large Area Telescope: a new light on the high-energy Universe}

\ShortTitle{The first year of the Fermi Large Area Telescope}

\author{\speaker{Luigi Tibaldo}\\
	Istituto Nazionale di Fisica Nucleare, Sezione di Padova, I-35131 Padova, Italy\\
	Dipartimento di Fisica ``G. Galilei'', Universit\`a di Padova, I-35131 Padova, Italy\\
	Laboratoire AIM, CEA-IRFU/CNRS/Universit\'e Paris Diderot, Service d'Astrophysique, CEA Saclay, 91191 Gif sur Yvette, France\\
        E-mail: \email{luigi.tibaldo@pd.infn.it}}

\author{on behalf of the \textit{Fermi} LAT Collaboration\\
       }

\abstract{For almost one year the Large Area Telescope on board the \textit{Fermi} observatory has been surveying high-energy phenomena in our Universe. We will present an overview of the status of the mission and of some results from the first year of observations, focusing on the topics of particular interest for the high-energy Physics community: detection of high-energy $\gamma$-ray bursts, the discovery of new populations of $\gamma$-ray sources, non-confirmation of the excess of diffuse GeV $\gamma$-ray emission seen by EGRET and, in greater detail, the recent measurement of the cosmic-ray electron spectrum from 20 GeV to 1 TeV.}

\FullConference{The 2009 Europhysics Conference on High Energy Physics,\\
		 July 16 - 22 2009\\
		 Krakow, Poland}

\begin{document}

Our Universe is home to violent phenomena, which can reach energies much higher than the collisions we produce in our particle accelerators. Therefore, the observations of high-energy astrophysical phenomena can provide deep insights on fundamental Physics, as well as information for understanding how our Universe works. In the last decades we expanded our investigation to a wide range of messengers, from charged cosmic rays to gravitational waves, from neutrinos to $\gamma$-rays. High-energy $\gamma$-rays play a key role, because they are relatively easy to detect and they have low interaction probabilities during their propagation.

\section{The \textit{Fermi} Large Area Telescope}\label{instrument}
The \textit{Fermi} Gamma-ray Space Telescope (formerly known as Gamma-ray Large Area Space Telescope, GLAST) was launched from the Cape Canaveral Air Station on June 11 2008 and it is on a nearly circular orbit around the Earth at 565 km of altitude with an inclination of $25.6^\circ$. The \textit{Fermi} observatory carries two instruments. The main instrument is the Large Area Telescope, hereafter LAT, a $\gamma$-ray imaging telescope detecting photons from 20 MeV to more than 300 GeV, with a field of view of 20\% of the sky at any instant and observing, in the normal sky-survey mode, the whole sky every $\sim 3$ hours. The secondary instrument is the Gamma-ray Burst Monitor (GBM), which supports the LAT in the observations of transient phenomena in the energy range from 8 keV to 40 MeV.

The \textit{Fermi} LAT is a pair-tracking telescope \cite{latpaper}, composed by an array of $4 \times 4$ towers. Each tower is made from a tracker (TKR) plus a calorimeter (CAL) module. The TKR has 18 $(x,y)$ layers of Silicon microstrip detectors interleaved with Tungsten foils to promote the $\gamma$-ray conversion into electron-positron pairs (12 thin foils of 0.03 radiation lengths in the front section plus 4 thick foils of 0.18 radiation lengths in the back section, the last two layers have no conversion foils), for a total depth of $\sim 1.5$ radiation lengths. The TKR is followed by a hodoscopic CsI CAL with imaging capabilities with a depth of $\sim 8.5$ radiation lengths. The whole system is surrounded by an anticoincidence detector (ACD), a segmented scintillator shield to discriminate the cosmic-ray background. On the bottom of the detector the Data Acquisition System is located. The LAT design and analysis \cite{latpaper} provide a sensitivity more than one order of magnitude higher than previous $\gamma$-ray telescopes, like EGRET on the \emph{Compton} Gamma-ray Observatory, and a superior angular resolution (e.g. the 68\% containment angle for front-converting events reaches $0.6^\circ$ at 1 GeV with respect to $1.7^\circ$ for EGRET). Therefore, the LAT is providing an unprecedented view of the gamma-ray sky.

\section{Highlights of Fermi $\gamma$-ray Science}
In this section we present a few highlights of the \textit{Fermi} $\gamma$-ray Science, selected on consideration of the interests of the audience and the personal taste of the speaker among the wide variety of results achieved during the first year of observations.

\begin{figure}
\begin{center}
\includegraphics[width=0.5\textwidth]{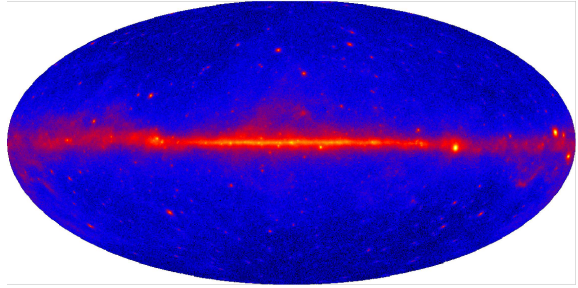}
\caption{Skymap obtained from the first 9 months of \textit{Fermi} observations (Galactic coordinates in Aitoff projection).
}\label{skymap}
\end{center}
\end{figure}

\begin{figure}
\begin{center}
\begin{tabular}{cc}
\includegraphics[width=0.48\textwidth]{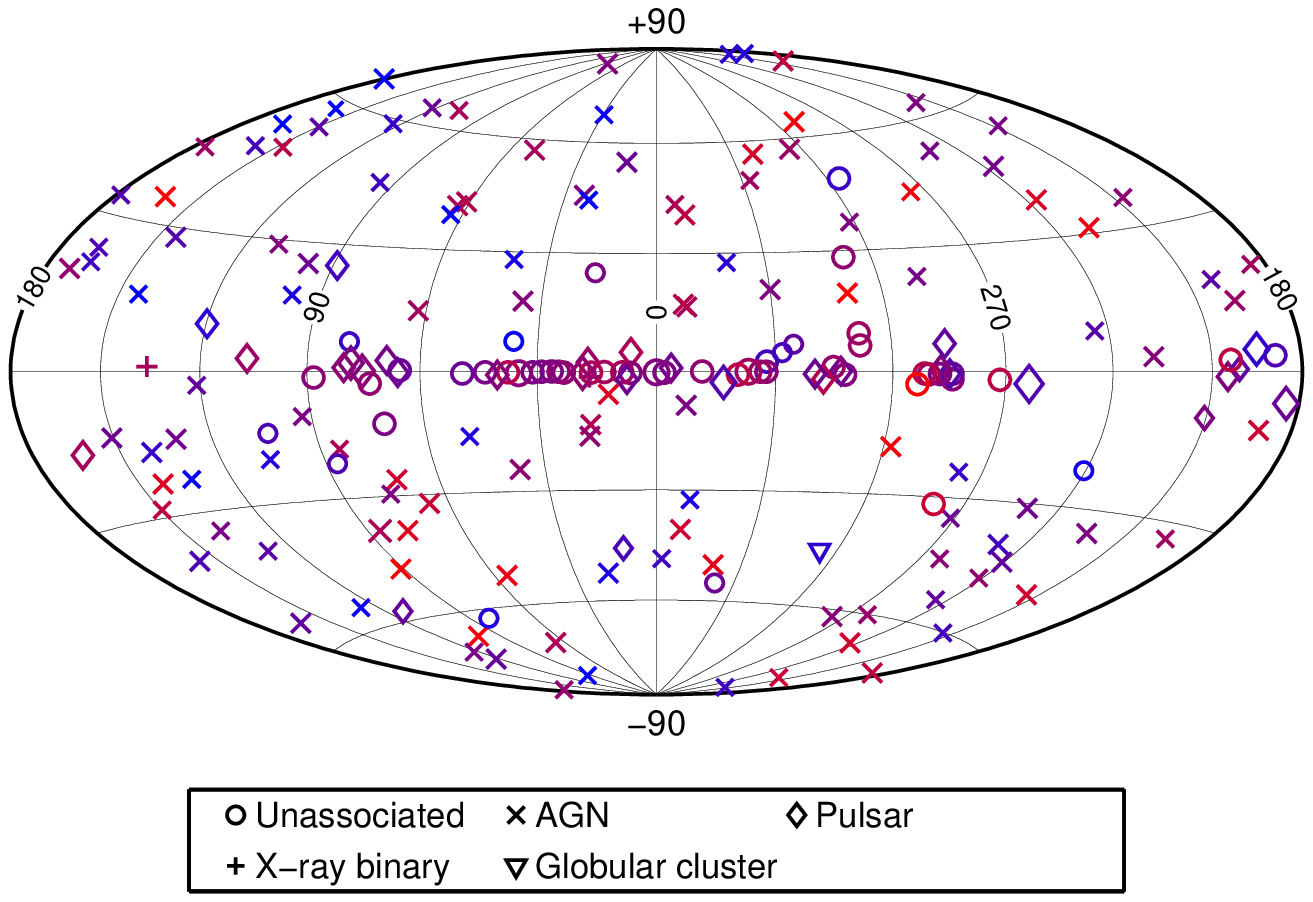}&
\includegraphics[width=0.3\textwidth]{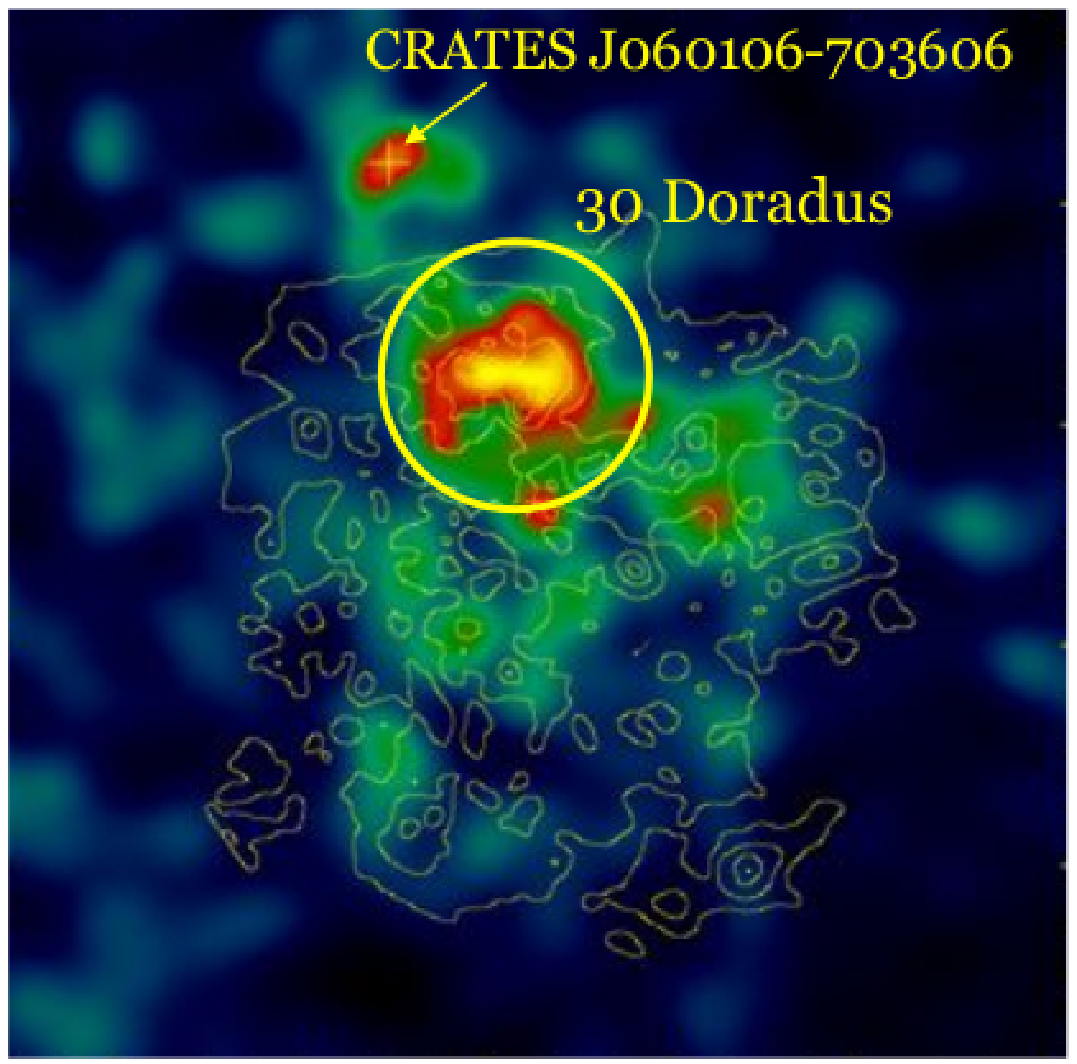}\\
\end{tabular}
\caption{On the left: position of the sources in the LAT Bright Source List \cite{BSlist} (Galactic coordinates in Aitoff projection). The colour represents the spectral type of each source, from softer (red) to harder (blue). On the right: (PRELIMINARY) skymap of LAT observations in the region of the Large Magellanic Cloud \cite{lmc}.} \label{resolution}
\end{center}
\end{figure}

The skymap from the first 9 months of observations in shown in Fig.~\ref{skymap}.
The LAT is resolving the $\gamma$-ray sky, allowing the detection of hundreds of point sources and permitting deeper understanding of the $\gamma$-ray diffuse emission. A major step toward the publication of the First Year Catalog has been the Bright Source List \cite{BSlist}, the list of the 205 sources detected with a significance $>10\sigma$ during the first three months of observations. The sources in the list are shown in Fig.~\ref{resolution}~(left). The LAT is able to resolve for the first time extended emission from extragalactic objects in high-energy $\gamma$-rays. In Fig.~\ref{resolution}~(right) we show a skymap of the region of the Large Ma\-gellanic Cloud (LMC), an external Galaxy where the LAT is resolving the $\gamma$-ray emission from the central starbust region 30 Doradus and the diffuse emission from the interactions of cosmic rays with the surrounding gas \cite{lmc}.

A characteristic feature of the $\gamma$-ray sky is its high variability. The LAT survey strategy is providing a coverage of the whole sky every three hours, thus allowing the monitoring of transient phenomena, like the flares of active galactic nuclei and $\gamma$-ray bursts. The first bright burst observed by the LAT was GRB 080916C \cite{brightgrb}, triggered by the GBM, at a redshift $z=4.35 \pm 0.15$ \cite{grond}. The LAT has detected more than 100 events above 100 MeV, up to $\sim 13$ GeV. The lightcurve of the burst shows an onset between low-energy and high-energy emission. Under the hypothesis that this
delay is an upper limit to the effect of the breaking of Lorenz invariance, a lower limit for the scale of quantum gravity
was derived, $M_\mathrm{QG}>1.3 \times 10^{18}$ GeV, just one order of magnitude below the Planck scale \cite{brightgrb}.

Among the wide variety of $\gamma$-ray sources, pulsars were the first objects identified in our Galaxy. However, in the EGRET era only six $\gamma$-ray pulsars were detected with high confidence, although $\sim 1000$ were known in the radio. Five of them were normal radio pulsars, discovered thanks to the ephemerides measured by radio astronomers. The last one, Geminga, was a radio-quiet pulsar identified after $\sim 20$ years of multiwavelength observations \cite{bignami}. The LAT has already unveiled 16 new pulsars, identified in blind searches using only $\gamma$-ray data \cite{blindpsr}. The LAT discovered also a new population of $\gamma$-ray pulsars with millisecond periods, known from radio observations as the second life of normal pulsars in binary systems and never observed before at high energies. The LAT detected strong $\gamma$-ray pulsations from 8 such ms pulsars \cite{mspsr}.

A phenomenon of great interest for particle physics, astrophysics and cosmology is the diffuse $\gamma$-ray emission. The Galactic diffuse emission (GDE) is produced by the interactions of cosmic rays (CRs) with the interstellar gas (via $\pi^0$ decay and Bremsstrahlung) and with the interstellar radiation field (via Inverse Compton, IC, scattering). The spectrum of the GDE measured by EGRET showed an excess of a factor $\sim 2$ in all directions on the sky at energies $\gtrsim 1$ GeV with respect to conventional models based on the locally measured CR spectra \cite{hunter,strong04}, interpreted by some authors as the long-awaited signature of dark matter. The first target of study for the LAT is the emission at intermediate Galactic latitudes, where most of the emission comes from interactions of CRs with the local gas.
\begin{figure}
\begin{center}
\begin{tabular}{cc}
\includegraphics[width=0.4\textwidth]{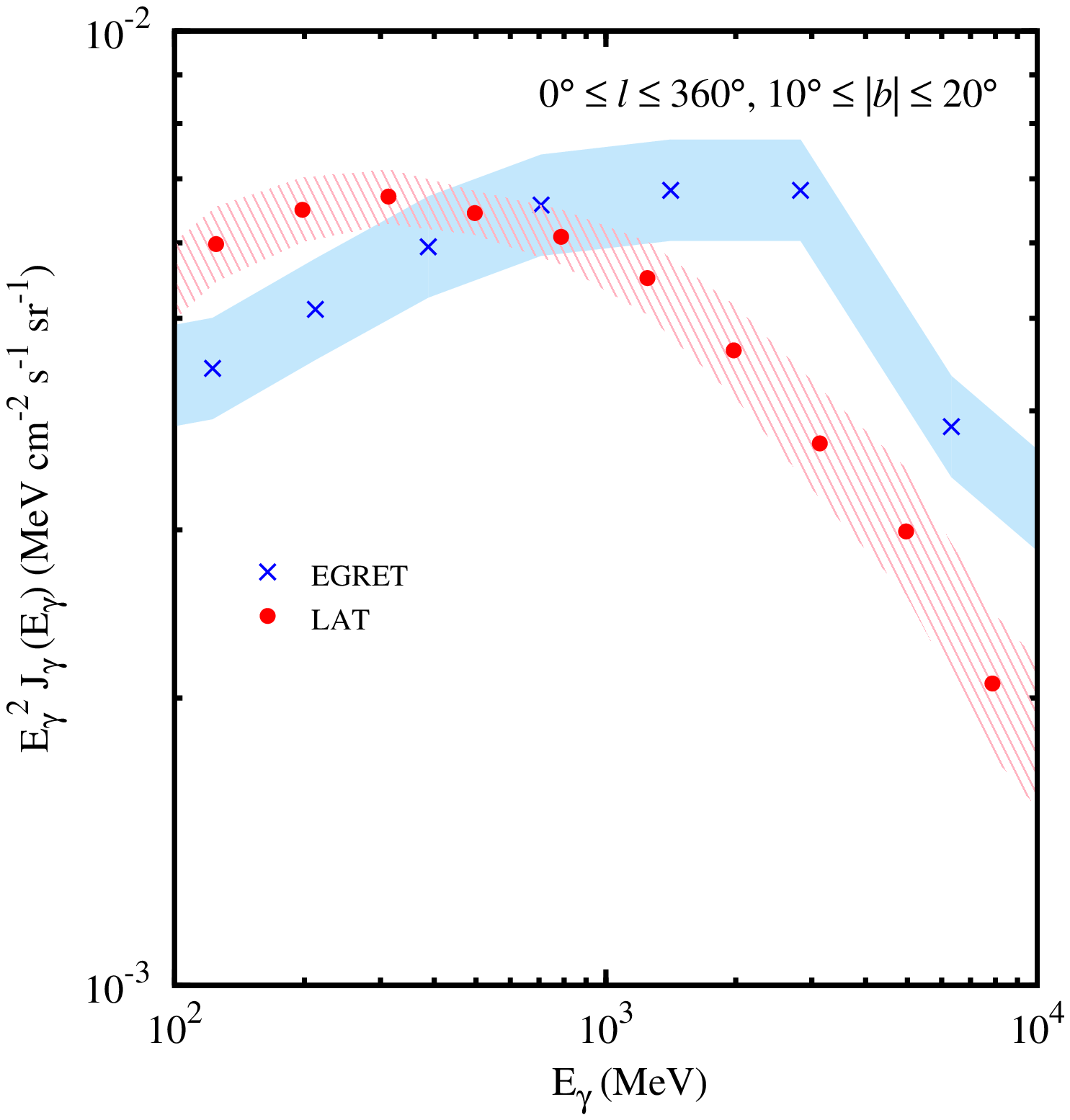}&
\includegraphics[width=0.4\textwidth]{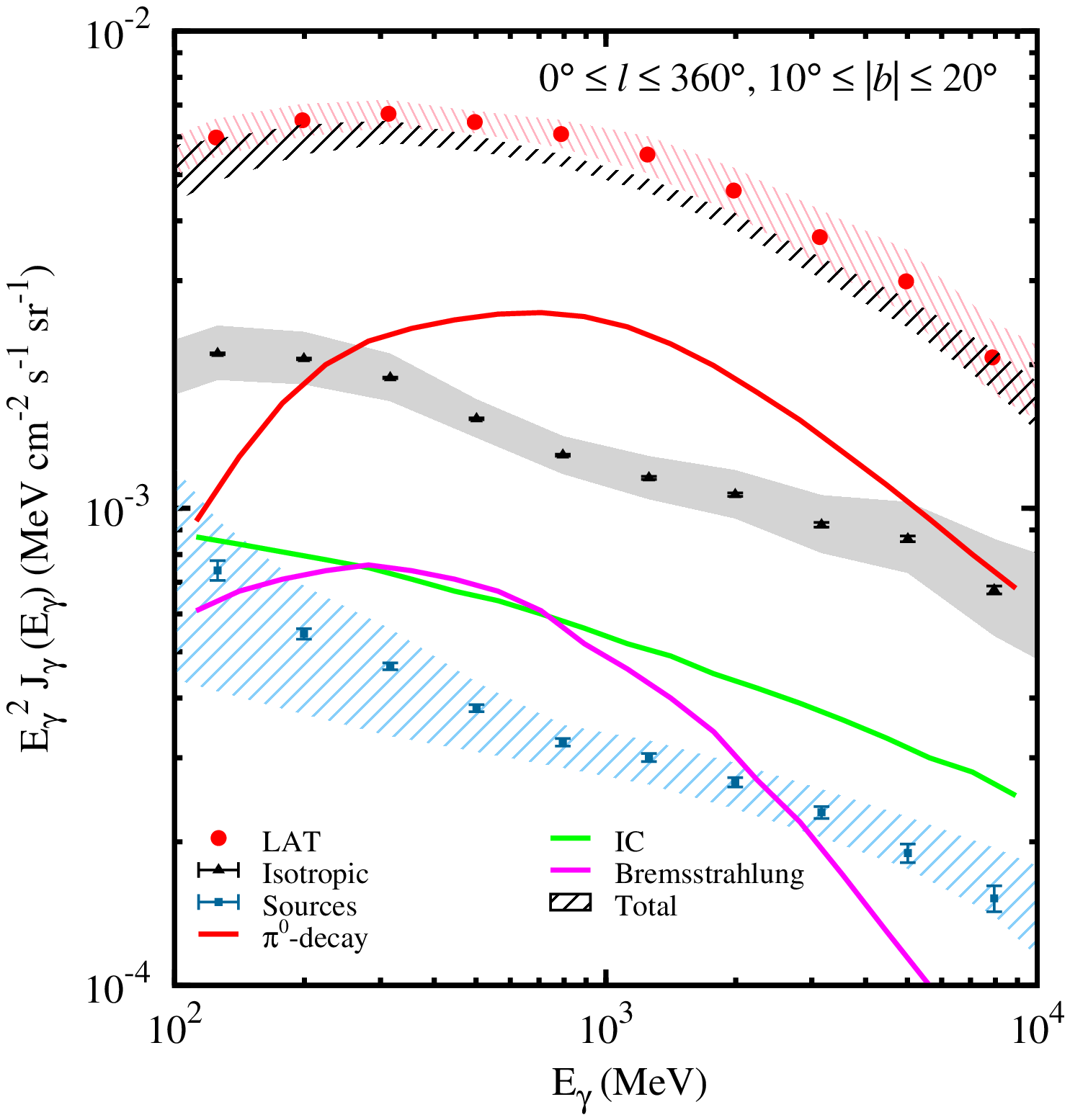}\\
\end{tabular}
\caption{PRELIMINARY -- Spectrum of the $\gamma$-ray emission measured by the LAT for energies between 100 MeV and 10 GeV, averaged over all longitudes in the region with Galactic latitude $10^\circ \leq |b| \leq 20^\circ$. Left: the measurement by the LAT (red) compared with the measurement by EGRET (blue). Right: comparison with the \emph{a priori} model for the GDE described in the text \cite{nongevexc}.} \label{ngefig}
\end{center}
\end{figure}
In Fig.~\ref{ngefig} (left) the spectrum measured by EGRET is compared with the spectrum measured by the LAT which is significantly softer. In Fig.~\ref{ngefig} (right) the same spectrum measured by the LAT is compared with an \emph{a priori} model for GDE emission, a revised version of the conventional model based on GALPROP \cite{strong04}, plus a point-source contribution and an isotropic unidentified background derived from fitting LAT data with the GDE model fixed. The LAT spectrum for this region of the sky is approximately reproduced by the GDE model based on the CR spectrum measured at Earth, thus being inconsistent with the GeV excess \cite{nongevexc}.

\section{Measurement of the cosmic-ray electron spectrum from 20 GeV to 1 TeV}
Designed as a highly-efficient $\gamma$-ray telescope (see Sec.~\ref{instrument}), the LAT is by its nature also a detector of electrons and positrons, because electromagnetic showers are germane to both interactions of electrons and photons in matter. The highly granular TKR and CAL and the hermetic, segmented ACD offer the opportunity to distinguish electrons from photons and hadrons. The large field of view of $\sim 2.4$ sr, the high efficiency and the on-board filter, configured to accept all the events which deposit more than 20 GeV in the calorimeter, provide to the LAT a large collecting power for high-energy electrons. However, the instrument does not have the capability of separating $e^+$ from $e^-$, so we refer to electrons as the sum of the two.

The main steps to select candidate electrons are \cite{eLAT}: 1 -- require events to fail the ACD vetoes developed to select $\gamma$-rays (providing an over-all contamination of photons $<2\%$ in the final electron sample); 2 -- apply several simple selection criteria relying on the capability to discriminate electromagnetic and hadronic showers based on their development in the TKR and the CAL (rejection power for hadrons of the order of $1:10^2$); 3 -- boost the selection using classification trees (enhancing the rejection power up to $1:10^3-10^4$).

\begin{figure}
\begin{center}
\includegraphics[width=0.5\textwidth]{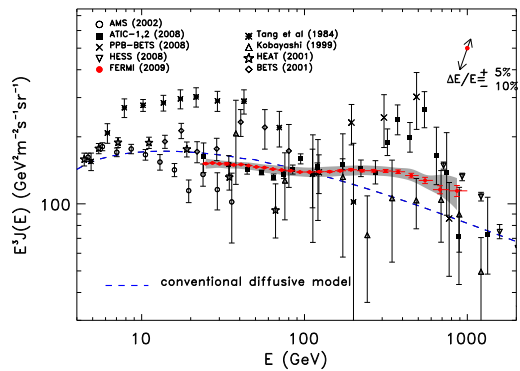}
\caption{The electron spectrum measured by the LAT (filled red circles). The systematic uncertainties on the event selection efficiency and the subtraction of the residual hadron contamination ($\lesssim 20\%$) are represented by the gray band. The two-headed arrow in the top-right corner represents the effect of systematic uncertainties on the absolute energy, constrained between $+5\%$ and $-10\%$ by the LAT beam test. Other measurements (black) are shown together with a conventional diffusive model \cite{strong04} (blue dashed line).}\label{espectrum}
\end{center}
\end{figure}

Then, the residual contamination by hadrons is evaluated using Montecarlo simulations and subtracted and the spectrum is corrected for the finite energy redistribution with an unfolding technique. The final spectrum is shown in Fig.~\ref{espectrum}.
LAT data are in good agreement with the measurement by HESS at higher energies \cite{hess}, but do not exhibit prominent spectral features as recently reported by ATIC \cite{atic}. If compared with a conventional model of CR propagation based on the GALPROP code \cite{strong04}, LAT electrons show a spectrum harder than expected. This spectrum, combined with the measurement by PAMELA of an increase in the ratio $e^+/(e^++e^-)$ up to 100 GeV \cite{pamela}, is very hard to reconcile with such conventional models. A natural explanation can be an additional local source of electrons, which might be given by astrophysical accelerators, like pulsars, as well as by dark matter particles. Further investigation is needed, and the LAT, with the studies of electrons as well as of $\gamma$-rays (diffuse emission, pulsars, dark matter searches \ldots), will play a fundamental role.

\acknowledgments
The \textit{Fermi} LAT Collaboration acknowledges support from a number of agencies and institutes for both development and the operation of the LAT as well as scientific data analysis. These include NASA and DOE in the United States, CEA/Irfu and IN2P3/CNRS in France, ASI and INFN in Italy, MEXT, KEK, and JAXA in Japan, and the K.~A.~Wallenberg Foundation, the Swedish Research Council and the National Space Board in Sweden. Additional support from INAF in Italy and CNES in France for science analysis during the operations phase is also gratefully acknowledged.


\begin{thebibliography}{99}

\bibitem{brightgrb} A.~A.~Abdo, et al. \emph{Fermi Observations of High-Energy Gamma-Ray Emission from GRB 080916C}, \emph{Science} {\bf 323} (1688)

\bibitem{eLAT} A.~A.~Abdo, et al. \emph{Measurement of the Cosmic Ray $e^++e^-$ spectrum from 20 GeV to 1 TeV with the Fermi Large Area Telescope}, \emph{Phys. Rev. Lett.} {\bf 102} (181101)

\bibitem{BSlist} A.~A.~Abdo, et al. \emph{Fermi/Large Area Telescope Bright Gamma-Ray Source List}, \emph{ApJS} {\bf 183} (46)

\bibitem{blindpsr} A.~A.~Abdo, et al. \emph{Detection of 16 Gamma-Ray Pulsars Through Blind Frequency Searches Using the Fermi LAT}, \emph{Science} {\bf 325} (840)

\bibitem{mspsr} A.~A.~Abdo, et al. \emph{A Population of Gamma-Ray Millisecond Pulsars Seen with the Fermi Large Area Telescope},  \emph{Science} {\bf 325} (848)

\bibitem{nongevexc} A.~A.~Abdo, et al. \emph{\textit{Fermi} LAT Measurements of
the Diffuse Gamma-Ray Emission at Intermediate Galactic Latitudes}, submitted

\bibitem{pamela} O.~Adriani, et al. \emph{An anomalous positron abundance in cosmic rays with energies 1.5 -- 100 GeV}, \emph{Nature} {\bf 458} (607)

\bibitem{hess} F.~Aharonian, et al. \emph{Energy Spectrum of Cosmic-Ray Electrons at TeV Energies}, \emph{Phys. Rev. Lett.} {\bf 101} (261104)
 
\bibitem{latpaper} W.~B.~Atwood, et al. \emph{The Large Area Telescope on the Fermi Gamma-Ray Space Telescope Mission}, \emph{ApJ} {\bf 697} (1071)

\bibitem{bignami} G.~F.~Bignami, P.~A.~Caraveo, \emph{Geminga: new period, old $\gamma$-rays}, \emph{Nature} {\bf 357} (287)

\bibitem{atic} J.~Chang, \emph{An excess of cosmic ray electrons at energies of 300–800 GeV}, \emph{Nature} {\bf 456} (362)

\bibitem{grond} J.~Greiner, et al. \emph{The redshift and afterglow of the extremely energetic gamma-ray burst GRB~080916C}, \emph{A\&A} {\bf 498} (89)

\bibitem{hunter} S.~D.~Hunter, et al. \emph{EGRET Observations of the Diffuse Gamma-Ray Emission from the Galactic Plane}, \emph{ApJ} {\bf 481} (205)

\bibitem{lmc} T.~A.~Porter, J.~Kn\"odlseder, et al. \emph{Fermi LAT Measurements
of the Gamma-Ray Emission from the Large Magellanic Cloud}, Proceedings of the
\emph{31$^{st}$ ICRC} {\tt arXiv:0907.0293v1}

\bibitem{strong04} A.~W.~Strong, I.~V.~Moskalenko, and O.~Reimer, \emph{Diffuse Galactic Continuum Gamma Rays: A Model Compatible with EGRET Data and Cosmic-Ray Measurements}, \emph{ApJ} {\bf 613} (962)

\end{thebibliography}
\end{document}